# Regional inflation analysis using social network data

Vasily Shcherbakov[1]      Ilia Karpov[2]

**Abstract.** Inflation is one of the most important macroeconomic indicators that have a great impact on the population of any country and region. Inflation is influenced by range of factors, one of which is inflation expectations. Many central banks take this factor into consideration while implementing monetary policy within the inflation targeting regime. Nowadays, a lot of people are active users of the Internet, especially social networks. There is a hypothesis that people search, read, and discuss mainly only those issues that are of particular interest to them. It is logical to assume that the dynamics of prices may also be in the focus of users' discussions. So, such discussions could be regarded as an alternative source of more rapid information about inflation expectations. This study is based on unstructured data from Vkontakte social network to analyze upward and downward inflationary trends (on the example of the Omsk region). The sample of more than 8.5 million posts was collected between January 2010 and May 2022. The authors used BERT neural networks to solve the problem. These models demonstrated better results than the benchmarks (e.g., logistic regression, decision tree classifier, etc.). It makes possible to define pro-inflationary and disinflationary types of keywords in different contexts and get their visualization with SHAP method. This analysis provides additional operational information about inflationary processes at the regional level The proposed approach can be scaled for other regions. At the same time the limitation of the work is the time and power costs for the initial training of similar models for all regions of Russia.

## Introduction

Information is a key element of the decision-making process in any field. In the economic sphere (especially at the macroeconomic level), the availability of timely and relevant information is of particular importance. Its status is especially enhanced during crises and turbulence.

---

[1] Vasily Shcherbakov, Ph.D., Head of the Economic Department, Omsk Region Branch of the Siberian Main Directorate of the Central Bank of the Russian Federation, shcherbakovvs@mail.ru

[2] Research Fellow at the International laboratory for Applied Network Research, Higher School of Economics, karpovilia@gmail.com

Almost all statistical data underlying macroeconomic decisions is published asynchronously and with certain lags. In economic theory and practice, this phenomenon has become widespread as a "jagged or ragged edge" (Giannone et al., 2008). This significantly complicates the decision-making on the implementation of public policy in the operational mode.

Nowadays many central banks all over the world, including the Bank of Russia, implement monetary policy within the inflation targeting regime. In this case monetary regulators set public quantitative inflation targets and pursue policy based on a range of other principles. For instance, the Bank of Russia makes its monetary policy decisions with help of the macroeconomic forecast and the analysis of a wide range of data[3].

As is known, inflation is influenced by a huge number of factors, one of which is inflation expectations (Frisch, 1990). It is regarded as a special factor measuring economic agents' assumptions regarding future inflation[4]. There is a large corpus of accumulated research in the field of perception of inflation expectations and price changes in general, both in the world and in Russian practice (Ranyard et al., 2008; De Bruine, 2011; Coibion et al., 2018; Gurov, 2022). For instance, it is shown that despite the information rigidities, there is a strong relationship between inflation and inflation expectations (Larsen et al., 2021).

So, on the one hand, implementing its monetary policy, the Bank of Russia takes into account and incorporates inflation expectations in its models and logics. On the other hand, it manages inflation expectations through explanation of its decisions and future intents to economic agents[5].

In Russia inflation expectations of households are measured based on InFOM survey findings on a monthly basis[6]. It should be noticed that the survey-based approach is generally accepted worldwide (for instance, Armantier et al., 2015; Schembri, 2020).

Despite the range of crucial advantages of this approach, it has a number of features which should be taken into account. So, *"inflation expectations measured based on household surveys almost always exceed actual inflation rates both in Russia and abroad. This difference is ascribed to the peculiarities of perception: people tend to notice and actively respond to price growth, whereas declining or stable prices usually attract less attention. Accordingly, people estimate inflation guided primarily by the product prices that have increased most significantly. Despite this systematic bias in the absolute values of inflation expectations, their change and relative level compared to the historical range are essential indicators showing possible changes in households' economic behavior. These changes in turn*

---

[3] Monetary policy. Bank of Russia. Retrieved from: https://cbr.ru/statistics/ddkp/objective_and_principles/ (Date of access: 12.12.2022)

[4] Monetary policy. Bank of Russia. Retrieved from: http://www.cbr.ru/eng/dkp/about_inflation/ (Date of access: 12.12.2022)

[5] Basically, changes of the key rate as the main instrument of the Bank of Russia's monetary policy

[6] Monetary policy. Bank of Russia. Retrieved from: http://www.cbr.ru/eng/analytics/dkp/inflation_expectations/#highlight=inflation%7Cexpectations (Date of access: 12.12.2022)

*influence future steady inflation"*[7]. Besides, the collection and processing of survey results takes some time and becomes available, as a rule, only on a monthly basis. In this regard, several areas can be identified for development in the field of assessment and analysis of inflation expectations, and their relationship with inflation.

Firstly, today more and more machine learning and deep learning methods are being adapted for use in the economic sphere, including for the purposes of analysis and forecasting of inflation (Chakraborty, Joseph, 2017; Baybuza, 2018; Pavlov, 2020; Peirano et al., 2021; Mamedli, Shibitov, 2021; Semiturkin, Shevelev, 2022). Also, there are also some papers aimed at inflation expectations and public opinion analysis at whole with help of these methods (Petkevič, 2018; Hu, 2019).

Secondly, the issue of finding alternative data sources, the use of which provides rapid information and gives new knowledge about the object under study, remains topical. For example, there is research that focuses on how consumers react to information provided by the media in concern of their inflation expectation formation. In this case the amount of news matters as much as the tone of news reports. At the same time, due to the noise level of data, agents experience significant extraction problems (Lamla, Lein, 2014). In this case different unstructured information (textual data form social networks, news and etc.) and search queries from Google, Yandex and other engines are applied for forecasting and nowcasting economic indicators, including inflation expectations (Goloshchapova, Andreev, 2017; Shcherbakov et al., 2022; Petrova, 2022).

Today, the majority of people are users of the Internet. In turn, social networks are the significant space for such discussions. *There is a hypothesis that people search and discuss on the Internet without coercion those issues that are of particular interest to them*. This hypothesis stays true for social networks as well.

It is logical to assume that the dynamics of prices of consumer goods and services may also be in the focus of users' discussions. In this case their search and discussion activities connected with the focus theme can be regarded as a proxy for inflation expectations. In other words, the study of posts published online in social networks can serve as additional information for the analysis of regional inflation processes. In this case, it becomes extremely important to filter a large amount of information and identify pro-inflationary and disinflationary keywords, statistics on which can be used for nowcasting inflation, obtaining leading indicators. So, the approach proposed in this paper is aimed at solving the problem with the help of machine learning methods

In this case if we are talking about such unstructured data as textual data from different social networks, it is quite reasonable to name two research: the paper by D. Aromi and M. Llada (Aromi, Llada, 2020) and the work by C. Angelico and colleagues (Angelico et al., 2022).

In fact, both works use data from Twitter as a proxy indicator for measuring inflation expectations. In the first mentioned paper the authors build an indicator of

---

[7] Inflation expectations and consumer behavior. No. 3 (75). March 2023. Bank of Russia. Retrieved from: http://www.cbr.ru/Collection/Collection/File/43865/Infl_exp_23-03_e.pdf (Date of access: 25.03.2023)

attention to inflation based on the corpus of Argentine tweets. The main idea is to compute the relative frequency of the noun "inflation" and the adjective "inflationary" in tweets. Then they use this indicator as one of the explanatory variables for inflation along with inflation rates for the previous periods and exchange rates. At the end the application of this indicator has improved the predictive power of their model for inflation (Aromi, Llada, 2020)

In this regard, the methods used in the second work are more advanced. C. Angelico and colleagues work with the corpus of Italian tweets. In order to filter the data and to lower the noise they apply a three-steps procedure, including a topic analysis with help of the method of Latent Dirichlet Allocation (LDA) and dictionary-based approach. As a result, the authors build directional Twitter-based inflation expectations indicators which provide additional information about inflation processes in Italy (Angelico et al.,2022).

In our point of view, the only soft point of the last paper is that at the very beginning the authors expertly determine the list of keywords which may be related to inflation (for example, oil prices or rents). Then they apply such a filter to all tweets. In other words, they predetermine the core of proxy indicators by themselves. With a high probability in this case there can be other meaningful words, which were not included in their list. In this paper we would like to overcome this issue.

In Russian reality, Twitter is not the most widespread social network, especially in today's conditions. Thus, in contrast to the above studies, in this paper the data from such social network as Vkontakte (or VK for short) is used. According to Similarweb[8] service VK takes the 21st position as one of the most visited websites in the World and the 4th position in Russia. So, this is the most popular social network in Russia.

We are going to focus on unstructured data from VK for the analysis of inflation in the Russian Federation, to be more precise – in the Omsk region. There are several reasons why the Omsk region was chosen as the research base. Firstly, the main emphasis was placed on testing the proposed methodology, which, in case of positive results, could be scaled relatively easily for other regions. A significant part of the study was devoted to data collection, data processing and the construction of appropriate models. The application of the considered approach to all regions at once without prior testing could lead to an irrational use of time resources and the need to use significantly more computing power. Secondly, there are more than 1.8 million VK accounts in the Omsk region. In this case the Omsk region can be considered as a representative region for Russia. According to official statistics, the share of the population using the Internet in Russia in 2021 was 90.1%, in the Omsk region - 90%[9]. Thirdly, one of the authors of this work permanently resides in the Omsk region and can verify the obtained data and results from the inside in terms of their qualitative values.

It should be noted that some authors have already collected data (texts of news posts, comments on them and some other relevant information) in VK and analyzed

---

[8] Similarweb. Retrieved from: https://www.similarweb.com/ru/top-websites/ (Date of access: 25.11.2022)

[9] The region of Russia. Socio-economic indicators. Federal State Statistics Service. Retrieved from: https://rosstat.gov.ru/storage/mediabank/Region_Pokaz_2022.pdf (Date of access: 25.03.2023)

it with the help of econometric and machine learning methods. But unlike the methodology reflected in this work, they do it on the basis of a limited number of news communities (for example, RBC, RIA Novosti, Vesti and etc. – not more than 15 communities) and for the Russian Federation as a whole, not for regional level. In addition, we have not found in these works the use of neural networks to assess inflation expectations based on social media data (Petrova, 2022; Shulyak, 2022).

In our study, we want to use unstructured data generated by VK posts to analyze upward and downward inflationary trends. In fact, we will deal with the problem of text classification - *determination of the pro-inflationary or disinflationary nature of users' statements*. To solve this problem, we will focus on using such state-of-art machine learning models as BERT-based models. It should be emphasized that the peculiarity of the study is the fact that we conduct such an analysis not on country, but on regional data (on the example of the Omsk region).

So, in order to achieve this goal, firstly we will describe the stages of data collecting and preprocessing. Then we will do an overview of the benchmark machine learning algorithms used in this study and a more detailed theoretical analysis of BERT models. At the end, we share the main results of the work and discuss future steps in this area of interest.

## 1. Data collection and preparation

There are two main types of data for analysis purposes: so-called structured and unstructured data. Usually, structured data consists of well-defined information in tabular form or in the form of organized databases. At the same time, unstructured data is a collection of files or media with different formats. Due to their nature, they are not grouped or classified. We can say that unstructured data is raw information that needs to be further processed. At the same time, such information may contain hidden data that are useful for explaining the observed processes.

As it was mentioned in the introduction, our main hypothesis is that VK users read and discuss topics that are important to them, including changes in prices for various goods and services. So, the intensification of such publications and discussions could be regarded as proxy for inflation expectations. To be more precise, we used textual posts from focused VK groups as a foundation for unstructured data.

At the very beginning, we will present the general logics (see Fig. 1), which we follow to collect the necessary data, and then provide more detailed information step by step.

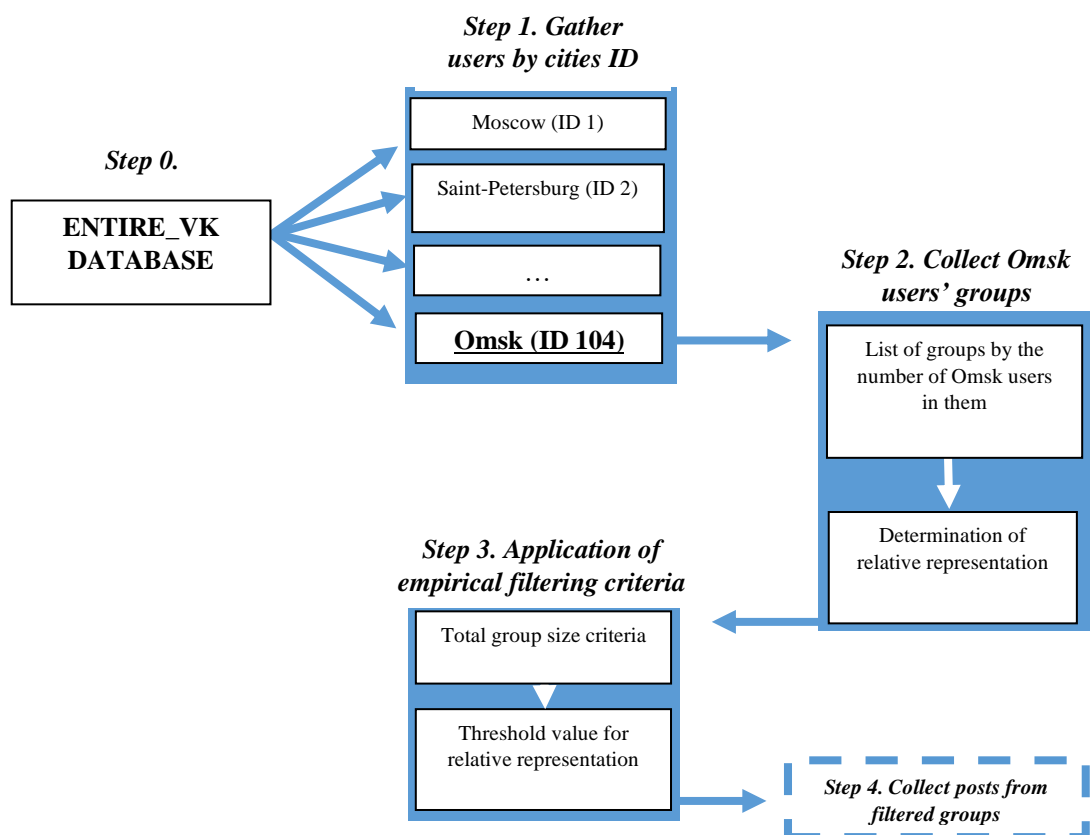

**Fig. 1.** General logic of data collection for the research
Source: prepared by the authors

At the first step, it was found out that there are users from 324 514 cities all over the World registered in VK and each of cities has an individual identifier. The Omsk region identifier is 104.

Then in the case of Omsk we got 1 839 190 users with such identifier. In other words, we found such an amount of VK users with Omsk roots. It should be noted that according to official statistics, 1 879 548 residents currently live in the Omsk region. It is reasonable to suppose that not all of them have VK accounts. Thus, there may be fake accounts and users who have more accounts than were collected by our code.

Also, it should be mentioned that it is possible that users registered in the Omsk region actually live in another region. Therefore, at the next step, we move on to the verification of Omsk groups, which could reflect what is happening in the region. In our opinion, with this approach, users who do not live in the region, but are interested in the events taking place in it (for example, through users' activities – comments, discussions, and reactions), can be fully used for analysis. We took these points into consideration.

We collected all VK groups they belong to. It turned out that there are 4 226 497 such groups. For analytical purposes, these groups were sorted in descending order of the number of Omsk residents in them.

In reality this approach has a very important meaning. There could be some groups which are registered as Omsk groups, but they might include a lot of users from other regions and/or bots, fake users. So, in this case posts that we are going to collect from filtered groups would not be relevant for our purpose.

Next it was necessary to find the total number of users from the groups which we had already collected. Due to the settings of some groups the total number of users for them was not found. Also, some groups were deleted. Thus, this fact can be considered as an additional filter for data collection.

At this step (*Step 3. Application of empirical filtering criteria*) we have introduced an additional hyperparameter in order to filter out our groups. This hyperparameter is responsible for the minimum number of users in the group. Empirically we chose it equal to 2000 users. By changing this hyperparameter in the range from 1500 to 2500, we did not observe a significant change in the groups (within 11%), so the average value of the examined range was chosen. In this case we can say that the indicator passed some kind of robustness test. On the one hand, information from many small groups do not bring any additional knowledge and can be considered as a kind of noise. On the other hand, using only large groups can lead to a potential underestimation of important information and data bias. It was necessary to find some tradeoff.

Applying this hyperparameter the total number of probably Omsk groups decreased from 4 226 497 to 2 229 812. We computed a relative representation of Omsk users in such groups and sorted them in descending order. In other words, we have calculated what percentage of users registered in the Omsk region are in these filtered groups. So, the logic is straightforward: the greater the proportion of Omsk users in groups, the more likely that topics (including regional inflation, price changes) are being discussed in them connected with processes occurring in the Omsk region.

This step brought us an interesting result. The highest relative representation of Omsk users was found in such group as "Дачник Омск"[10] – 56,82%. But we still had a lot of groups (more than 2.2 million), and they could have a lot of information noise. We decided to visualize these groups by relative representation value. At the beginning we build a standard scale chart (see Fig. 2).

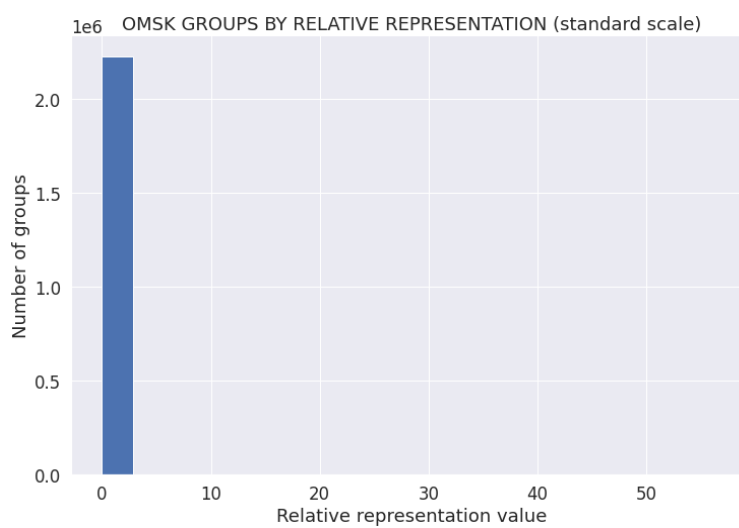

---

[10] VK community. Retrieved from: https://vk.com/public148267576 (Date of access: 01.07.2022)

**Fig. 2.** VK groups of Omsk origin by relative representation (standard scale)
Source: prepared by the authors

Visually, there is a bias in groups with a small percentage of Omsk residents. In other words, we do not observe any visible values on the right side of the tail. In such a way it is impossible to determine the threshold value for relative representation. Because of this, we have plotted the same data on a logarithmic scale (see Fig. 3).

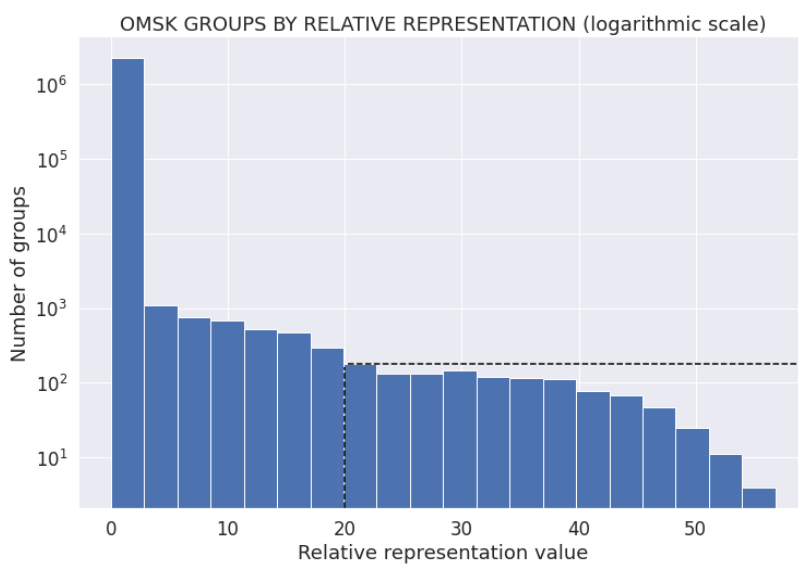

**Fig. 3.** VK groups of Omsk origin by relative representation (logarithmic scale)
Source: prepared by the authors

This plot provided more sufficient visualization. We still see a shift on the left side of the group distribution, but now we can determine the threshold value. Based on the analysis, we chose 20% as the minimum relative value for the filtered groups. We have highlighted this area with dotted lines (see. Fig. 3). In this case, the distribution of groups is shown in the following figure (see Fig. 4).

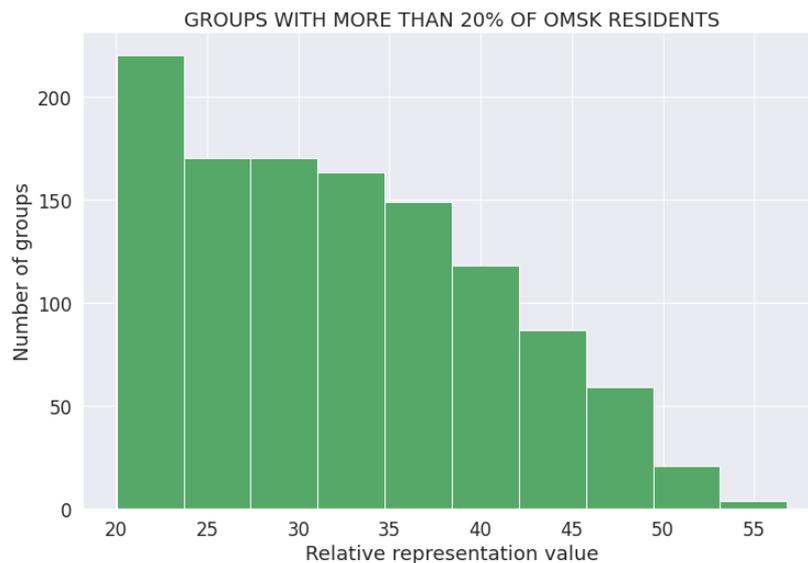

**Fig. 4.** Groups with more than 20% Omsk residents' distribution
Source: prepared by the authors

So, the target sample (the number of groups with Omsk users is more than 20% of the total number of group members) included 1 161 groups. We focused on them and collected the information that was posted there in the period from January 2010 to May 2022. As a result, we got the target data frame with 8 518 428 posts.

At the last stage, this data is labeled based on the dynamics of inflation in the Omsk region. Inflation data at the regional level, in contrast to the country level, is available only in monthly dimensions. The Federal State Statistics Service publishes weekly inflation data only for the Russian Federation as a whole. This fact is a significant limitation within the framework of the work.

As was mentioned before, we use unstructured data generated by VK posts to analyze upward and downward inflationary trends. So, our attention is focused on determination of the pro-inflationary or disinflationary nature of users' statements. Therefore, we labeled the inflation data based on an upward or downward trend of inflation. Hereinafter it allows us to transfer the problem to the field of binary classification.

After that we used such function as argrelextrema from scipy library for Python in order to calculate the relative extrema of our data (minimas and maximas). It was noticed that there were a lot of close extremes because of data specifics (volatility). Therefore, it was decided to eliminate some of them in neighborhoods (smoothing for 3-month period). As a result of additional analysis, we got 9 breaking points:

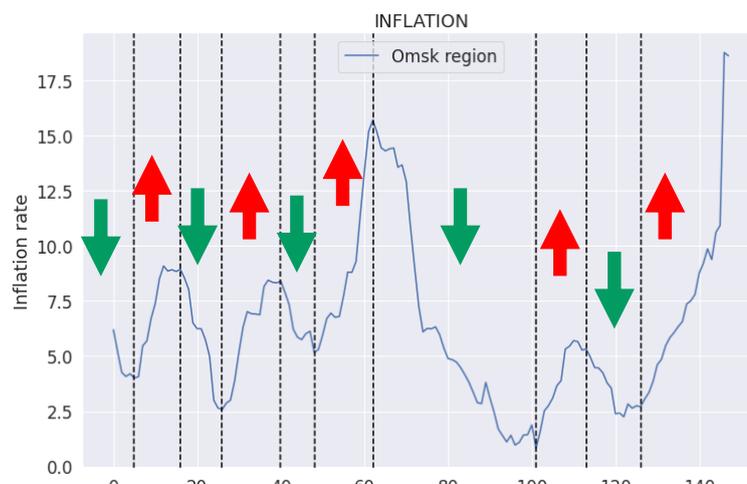

**Fig. 5.** Final relative extremes in inflation dynamics in the Omsk region from January 2010 to May 2022
Source: prepared by the author based on
https://www.fedstat.ru/indicator/31074 (Date of access: 26.08.2022)

So, the list of extremes is the following: {5: 'June 2010', 16: 'May 2011', 26: 'Mar. 2012', 40: 'May 2013', 48: 'Jan. 2014', 62: 'Mar. 2015', 101: 'June 2018', 113: 'June 2019', 126: 'July 2020'}

Based on these findings we labelled our dataset in two classes: "1" – increase in inflation and "0" – decrease in inflation. It should be emphasized that 43.6% of the data were marked as the 1st class representations and 56.4% as the 2nd class representations.

## 2. Methodology

The main focus of this study will be on application of BERT-based model (Bidirectional Encoder Representations from Transformers) for our classification task. This model was first proposed by Jacob Devlin, Ming-Wei Chang, Kenton Lee, and Kristina Toutanova. The result of their study was published in 2019 (Devlin et. al., 2019). Historically this model was applied for multilingual translation purposes by Google.

Since then, this approach has gained wide popularity and has been used to solve a variety of tasks, including classification ones. By now, the BERT family of models is the most used worldwide. So, according to the specialized portal Hugging Face, the most downloaded model of all existing is "bert-base-uncased". It has been applied more than 47 million times[11].

It should be noted that despite the fact that there are a large number of variations of models based on BERT, the fundamental idea remains the same. Therefore, in this part of the paper, we want to highlight the theoretical foundations of this approach, as well as describe a number of machine learning models that can be used for similar tasks. In this work, they will be applied as a benchmark for comparing the metrics of the BERT model.

It was emphasized before that we are going to solve binary classification problem in this study. The following indicators will be used as metrics for comparing

---

[11] Hugging Face. Retrieved from: https://huggingface.co/models?sort=downloads (Date of access: 01.04.2023)

models: precision, recall and f1-score. The formulas for calculating these indicators[12] are shown below:

$$precision = \frac{TP}{TP+FP} \quad (1)$$

$$recall = \frac{TP}{TP+FN} \quad (2)$$

$$f1-score = 2*\frac{precision*recall}{precision+recall} = \frac{TP}{TP+0.5*(FP+FN)} \quad (3)$$

Each of these metrics has its own limitations, so they are usually used in a complex. For instance, recall demonstrates the ability of the algorithm to detect a class in general, and precision demonstrates the ability to distinguish this class from other classes. In turn, the f1-score indicator acts as an aggregated indicator for precision and recall. As a result, more emphasis will be placed on f1-score.

**2.1. BERT-based models**

The issue of solving problems of natural language processing has an evolutionary nature. At various times, a lot of approaches have been provided for this purpose. For instance, so-called feed-forward neural network (Bengio et. al., 2003), various variations of recurrent neural network (RNN) (Mikolov et. al., 2010), and convolutional neural network (CNN) (Bradbury et. al., 2016) have been used and continue to be used for these purposes.

Nowadays the BERT-based models have got the status of state-of-art models. It is based on the idea of the Transformer (Vaswani et. al., 2017). It is known that recurrent neural network (RNN) process input data sequentially, convolutional neural network (CNN) process them in parallel but have access to only a few nearest words (set by the convolution size). In turn, the Transformer has access to all the words in the sequence and process them in parallel (Kuratov, 2020) due to its architecture.

The Transformer architecture consists of repeating fully connected layers and attention mechanisms forming the Transformer layers. In turn they make up the encoder and decoder sequences only due to the mechanism of attention, without recurrence and convolutions (Vaswani et. al., 2017; Kuratov, 2020). The reduced dimension meant that the total computational cost was similar to that of single-head attention with full dimensionality. The proposed approach has made a significant contribution to the development of the studied direction. The language models based on the Transformer demonstrated relatively better quality (for example, Liu et. al., 2019; Lan et. al., 2020).

BERT is not the only one model which uses the Transformer idea. For instance, there are such models as GPT and ELMo models. But bidirectionality distinguishes the BERT model from them. In GPT, only the left context is used to build token representations. There are representations from two independent recurrent networks in ELMo: one processes the sequence from left to right, the second from right to left. So, it also has bidirectionality, but the views from each direction are independent (Kuratov, 2020). Also, unlike the previous models, BERT is the unsupervised language representation, pre-trained using only a plain text corpus (in this case, Wikipedia) (Devlin, Chang, 2018).

---

[12] Here and further following the classical logic of confusion matrix we have such notations: TP – true positive, TN – true negative, FP – false positive, FN – false negative.

In this study we are going to use the BERT-based model in text classification task. The scientific interest is to determine whether the use of the BERT-based model will allow to extract new knowledge from an array of unstructured data, to be more precise – *could we get an additional information about pro-inflationary and disinflationary keywords in different contexts not expertly but based on big data analysis?* Special attention should be paid to the fact that such a model will be used to classify upward and downward inflation trends at the regional level (on the example of the Omsk region)

Large-scale pre-trained models, including BERT, have become a milestone in the field of artificial intelligence. They can effectively capture knowledge from massive labeled and unlabeled data. It is noticed that *"by storing knowledge into huge parameters and fine-tuning on specific tasks, the rich knowledge implicitly encoded in huge parameters can benefit a variety of downstream tasks, which has been extensively demonstrated via experimental verification and empirical analysis"* (Han et. al., 2021).

The use of pre-trained models became useful after the introduction of transfer learning approaches (Thrun, Pratt, 1998). Transfer learning let researches to use the experience gained in solving one problem to solve another, similar problem. So, the neural network is first trained on a large amount of data (pre-training), then on the target set (fine-tuning). In other words, this step allowed us to resolve new problems with relatively small samples of data. If we are talking about initial BERT model, there were 2 types of pre-trained models in 2 variants (cased and uncased)[13].

Nowadays a much larger number of models have been trained and made publicly available. There are some pre-trained BERT models for other languages. Therefore, we work with Russian-speaking unstructured data in this study, it is important for us to deal with such models. In this case we should mentioned RuBERT[14].

This model was build based on multilingual model from the BERT repository[15]. Training of the subword vocabulary was performed on the Russian part of Wikipedia and news data. So, BERT multilingual and RuBERT have the same size of vocabulary, but the second one's vocabulary was built especially for Russian language. During tests, for example, sentiment analysis of posts from VK, RuBERT showed better results than BERT multilingual, as well as logistic regression, gradient boosting, and other machine learning algorithms (Kuratov, Arkhipov, 2019).

---

[13] BERT. Retrieved from: https://github.com/google-research/bert#pre-trained-models (Date of access: 10.12.2022)
[14] DeepPavlov. Retrieved from: http://docs.deeppavlov.ai/en/master/features/models/bert.html (Date of access: 10.12.2022)
[15] 104 languages, 12-layer, 768-hidden, 12-heads, 110M parameters

RuBERT, along with other large BERT models, demonstrates its high efficiency. At the same time, they are computationally expensive. So, there is some kind of trade-off in this case – creation of small models. It is admitted that "*there are many aspects to explore: the parametric form of the compact model (architecture, number of parameters, trade-off between number of hidden layers and embedding size), the training data (size, distribution, presence or absence of labels, training objective) and etc.*" (Turc et. al., 2019).

In other words, small models are designed for environments with limited computational resources. These models could be fine-tuned in the same way as the original BERT models. However, they are most effective in the context of knowledge distillation, where the fine-tuning labels are created by a larger and more accurate teacher[16]. It has been confirmed that distillation shows good results for transferring knowledge from an ensemble or from a large highly regularized model to a smaller, distilled model (Hinton et. al., 2019).

In this regard, RuBert-tiny (small variant of RuBERT) was used as the primary model[17]. Without any doubt, there is a certain compromise between the computational cost, quality, and speed of this model. Nevertheless, it was decided to use this type of model to take the first steps in studying regional inflation based on unstructured data from VK groups.

We have adapted the approach[18] proposed by Shitkov K. In our case the batch size equal to 32 was used. The initial BERT model has a maximum length equal to 512. We experimented with maximum sentence length for padding / truncating to. It is known that the maximum length does impact training and evaluation speed. So, based on this hyperparameter, 4 variants of models (64, 128, 256 and 512) were created.

**2.2. Benchmark models**

Based on literature analysis logistic regression, decision tree classifier, random forest classifier and gradient boosting classifier were chosen as benchmark models for comparison with mentioned above variants of BERT-based model. To build the benchmark models, ready-made algorithms are used from scklearn[19] library for Python. It should be noted that each of the models has its own settings. Therefore, to tune the hyperparameters of the benchmark models, we use GridSearchCV from sklearn. This technique also allows us to apply cross-validation method to the data.

Logistic regression can be used to classify an observation into one of two classes, or into one of many classes (multinomial logistic regression). In our case we have two classes, and this method can be considered applicable. As a rule, logistic regression uses the so-called sigmoidal function. Such function is useful because it can take as an input any value (from negative infinity to positive infinity), whereas the output is limited to values from 0 to 1.

---

[16] BERT. Retrieved from: https://github.com/google-research/bert#pre-trained-models (Date of access: 10.12.2022)
[17] Rubert-tiny. Retrieved from: https://huggingface.co/cointegrated/rubert-tiny (Date of access: 09.08.2022)
[18] Bert for classification Retrieved from: https://github.com/shitkov/bert4classification (Date of access: 09.09.2022)
[19] Scikit-learn. Retrieved from: https://scikit-learn.org/ (Date of access: 25.11.2022)

Logistic regression is one of the simplest and most widely used methods in machine learning, including text classification area. For example, Shah K. and his colleagues used different models based on machine learning algorithms in order to build a BBC news text classification. The authors made the conclusion that logistic regression classifier demonstrated the highest accuracy for the data set. The second-best result was shown by the random forest classifier (Shah et. al., 2020).

Santosh Baboo S. and Amirthapriya M. also came to the conclusion that logistic regression outperformed other models (random forest, stochastic gradient boosting). They used these models for the Twitter posts classifications based on the variety of emotions (Santhosh Baboo, Amirthapriya, 2022).

Decision tree algorithm is also one of the widely used techniques in data mining systems that creates classifiers. There are different types of decision tree algorithms. The summary and analysis of such algorithms are given in the paper by Jijo B.T., Abdulazeez A. M. (Jijo, Abdulazeez, 2021). Generally, a decision tree classifier is a variant of supervised machine learning algorithm that predicts a target variable by learning simple decisions inferred from the sample's features. The decisions are all split into binary decisions (either a yes or a no) until a label is calculated[20].

Random forests is an ensemble learning algorithm for classification and other purposes. It operates by constructing a multitude of decision trees at training time. There is a special mechanism of randomness to decrease the variance of the forest estimator. It is found out that individual decision trees typically exhibit high variance and tend to overfit. Random forests achieve a decrease in variance by combining different trees, sometimes at the cost of a small increase in bias. In practice, the decrease in variance is often significant, which leads to an overall improvement in the model[21][22]. Some papers show that the use of improved random forest classifiers can demonstrate better result than other algorithms, including SVM and logistic regression, in the case of text classification (Jalal et. al., 2022).

Gradient boosting is also one of the ensembles machine learning methods. It iteratively learns from each of the weak learners to build a strong model. To be deeper an algorithm minimizes a loss function by iteratively choosing a function that points towards the negative gradient. Generally, it can deal with regression, classification, and ranking tasks[23].

Another important point is the preparation of data for benchmark models. These models cannot directly process textual information unlike BERT models. So, we need to provide text data vectorization in order to use it in the chosen machine learning models. It is known that in a large text corpus (as well as in our case) some words appear with higher frequency, but do not carry meaningful information. If we did not take this issue into account, those very common terms would shadow the frequencies of rarer yet more informative terms. In order to re-weight the count features into floating point values suitable for usage by a classifier, we use special techniques. One of the most common ways to solve this issue is to use TfidfVectorizer[24].

---

[20] Decision Tree Classifier. Retrieved from: https://scikit-learn.org/stable/modules/generated/sklearn.tree.DecisionTreeClassifier.html (Date of access: 27.11.2022)
[21] Ensemble method. Retrieved from: https://scikit-learn.org/stable/modules/ensemble.html#forest (Date of access: 27.11.2022)
[22] Random Forests. Retrieved from: https://datagy.io/sklearn-random-forests/ (Date of access: 27.11.2022)
[23] Gradient Boosting Classifier. Retrieved from: https://scikit-learn.org/stable/modules/generated/sklearn.ensemble.GradientBoostingClassifier.html (Date of access: 27.11.2022)

For instance, we see the application of this method in such spheres of text classification as: detecting spams and fake news (Ahmed et. al., 2017), Instagram caption classification (Ramadhani, Hadi, 2021) and many others.

The general logic is as follows: the simpler, the better. In this case if the BERT model beats the benchmark models even using the raw data, it can be interpreted as an additional signal for the development of this direction. And vice versa if, at the end, the BERT model, which is costly both in terms of the required computing power and in terms of time resources, will not be able to surpass the benchmark models, then the question will arise about its applicability and/or options for finding additional ways to optimize it.

## 3. Main results

In this part of the paper, we will show the main results, including the quality metrics of BERT-based models and benchmark models. At the first step we divided our data into 3 samples (train, validation, and test) in this proportion:
- Train sample takes up 80% of all data (5.1 million of observations)
- Validation and test sample are equal and take up 20% of all data (1.7 million of observations)

It should be additionally admitted that the validation sample was obtained from the train sample and amounted to 25% of it.

Now let us deal with BERT models modifications. We built the 4 variants of models based on maximum sentence length parameter (64, 128, 256 and 512). All models had the same batch size equal to 32 and were trained for 5 epochs. The approximate time to train these models using the GPU is shown in the following table 1:

**Table 1**
**Time to train models**

| # | Model modifications | Time to train (rounded) |
|---|---|---|
| 1 | BERT-inflation-64 | 19 hours |
| 2 | BERT-inflation-128 | 28 hours |
| 3 | BERT-inflation-256 | 54 hours |
| 4 | BERT-inflation-512 | 165 hours |

Source: prepared by the authors

---

[24] TfidfVectorizer. Retrieved from: https://scikit-learn.org/stable/modules/generated/sklearn.feature_extraction.text.TfidfVectorizer.html#sklearn.feature_extraction.text.TfidfVectorizer (Date of access: 27.11.2022)

In all models AdamW was used as the optimizer. It has improved weight decay in comparison with simple Adam. In this case weight decay can be regarded as a form of regularization to lower the chance of overfitting. The learning rate was equal to 2*10-5(2e-5).

We visualized training and validation losses by epochs (for all variations of the model). These loss curves provide a better insight into how the learning performance changes over the number of epochs. Also, it can help diagnose possible problems with learning that may lead to an underfit or an overfit model. Based on the figure below we can conclude that our models are quite close to a good fit because we can observe decreasing curves' types and also there is a small gap between validation and training losses. Further increase of epochs may lead to overfitting of the models. These visualizations are shown below (see Fig. 6)

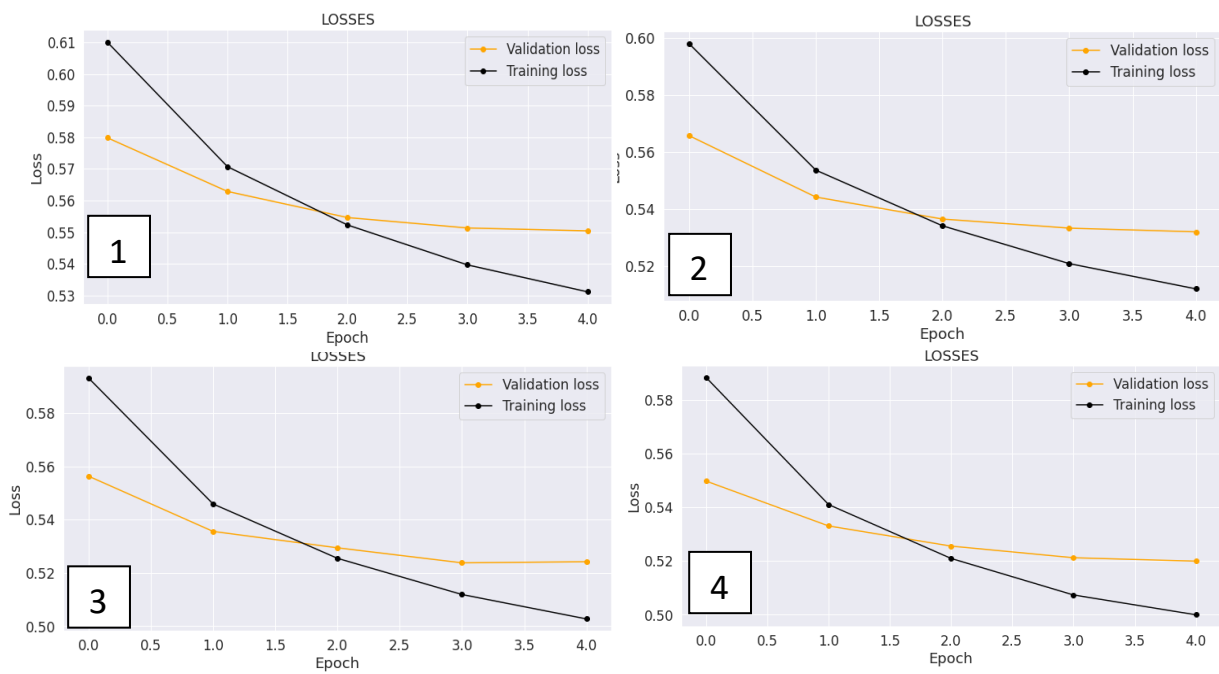

**Fig. 6.** Training and validation losses by epochs (1- Bert-inflation-64; 2- Bert-inflation-128; 3- Bert-inflation-256; 4- Bert-inflation-512;)
Source: prepared by the authors

As it was shown earlier, we used 4 machine learning algorithms (logistic regression, decision tree classifier, random forest classifier, and gradient boosting classifier) as the benchmark models. The comparative analysis of these models based on f1-score, precision and recall is shown in table 2.

**Table 2**
**Metrics by the models**

| # | Models | Recall | Precision | F1-score* |
|---|---|---|---|---|
| 1 | BERT-inflation-512 | 0.7027 | 0.7268 | 0.7050 |
| **2** | **BERT-inflation-256*** | **0.7007** | **0.7218** | **0.7030** |
| 3 | BERT-inflation-128 | 0.6960 | 0.7144 | 0.6982 |
| 4 | BERT-inflation-64 | 0.6816 | 0.7025 | 0.6831 |

| 5 | Logistic Regression (C=1.0, max_iter=1000) | 0.5516 | 0.5878 | 0.5198 |
| 6 | Gradient Boosting Classifier (learning_rate=0.05, n_estimators=200) | 0.5349 | 0.6343 | 0.4564 |
| 7 | Decision Tree Classifier (max_depth =10) | 0.5308 | 0.6162 | 0.4519 |
| 8 | Random Forest Classifier (max_depth =10) | 0.5273 | 0.6719 | 0.4297 |

Source: prepared by the authors

It should be recalled that in the framework of this work we solve the problem of binary text classification. Firstly, we trained the models on the training sample based on the VK posts. Secondly, we used these models to determine the nature of inflationary processes (acceleration or deceleration of inflation) in different time periods based on the test sample. So, we have obtained metrics for the quality of these models. This approach will further allow us to assess inflation expectations (sentiment) based on the information received from the social network in real time or with a slight delay. As a result, the received information can be used for nowcasting inflation in general.

As we can see, the BERT-based models show much better results even based on raw data without any cleaning procedures in comparison with classical machine learning models. Despite the slight superiority of the BERT-inflation-512 model compared to the BERT-inflation-256 model (based on f1-score), in our opinion, the second model is the most preferable, if we take into account the time spent on training on the collected data.

We will use this model modification for next steps connected with interpretability and discussion part. At the same time, it should be admitted that the best benchmark model is logistic regression (see table 2). It follows the logic of mentioned papers in this field.

## 4. Discussion

As it was rightly noted, regardless of the ultimate goal of someone's solutions in the field of data science, the final results are always preferable to interpret and understand. Without doubt, it will help to validate and improve these solutions[25]. In other words, the problem of the so-called "black box" stays true. There are many aspects related to the issue of interpretability of the model, including trust, causality, transferability, informativeness, ethical issues, and others (Lipton, 2016).

Nevertheless, realizing the importance of this issue, we have taken several steps in this direction. We used SHAP (SHapley Additive exPlanations) in order to achieve this goal. As it is stated in the official documentation of this method, *"it is*

---

[25] Towards data science. Retrieved from: https://towardsdatascience.com/interpretability-in-machine-learning-70c30694a05f (Date of access: 27.03.2023)

*a game theoretic approach to explain the output of any machine learning model. It connects optimal credit allocation with local explanations using the classic Shapley values from game theory and their related extension*"[26].

One of the advantages of such methods is the ability to visualize the results. SHAP is not the only one method for model's interpretability (for instance, LIME and other). But there is a number of advances that have been made in this matter (Kokalj et. al., 2021; Subies et. al., 2021).

In our case, we have created a special function that could make visualization of the interpretation of posts from VK, or other texts based on our pretrained model. Just to remind in our analysis the BERT-inflation-256 was approved as the best model.

It should be admitted since we dealt with Russian-based posts from VK and used the special Russian-based BERT model (RuBert-tiny), we can provide the example of such visualization only in Russian language (see. Fig. 7). However, in our opinion, this point is not the key one, since the main idea is to determine the pro-inflationary (red) and/or disinflationary (blue) nature of the words contained in the posts. Regardless of the language used, the demonstrated results are illustrative examples of interpreting models using SHAP.

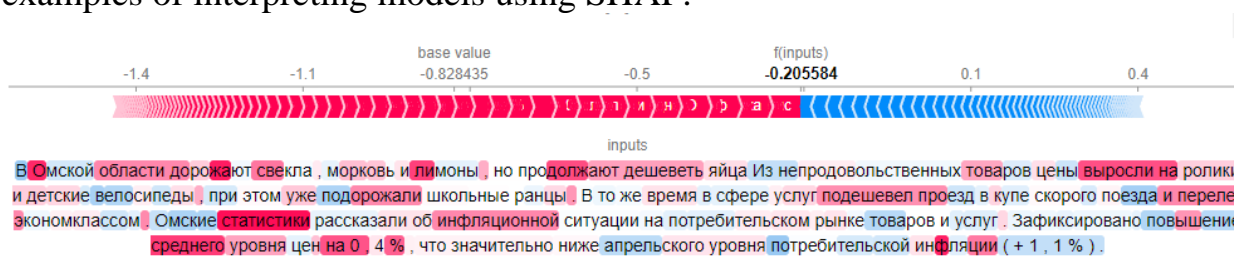

**Fig. 7**. Example of the model interpretation by SHAP method based on the selected posts from VK
Source: prepared by the authors

At the very beginning it was noticed that the majority of papers in this field use predetermined list of words by experts in order to value inflation expectations based on unstructured data from social networks or search engines. In this case there is probability that we can miss some meaningful keywords.

As shown in the example, the model quite clearly caught some of the marker words associated with price changes and inflation in general without any predetermination. For example, it admitted such words as "increase" ("выросли на"), "inflationary" ("инфляционной"), the "cost of building materials" ("стоимость стройматериалов") and others. The colors of words (red and blue) and their intensity say about pro-inflationary and disinflationary types of them. The applied method pays great attention to the context of use of different words. This fully fits into the general economic logic.

At the same time, certain difficulties with tokenization of a number of words are visible, which may be a further step to improve the described approach. So, this study allowed us to identify further areas for the development of this direction. Here we would like to highlight the main ones:
- Use of other methods of marking up data, for example, into three classes depending on the level of monthly inflation ("0" - up to 4% - low; "1" - from 4% to 8% - "medium"; "2" - over 8% - "high")
- Improvement of the methods to text tokenization for the Russian based BERT model
- Further adaptation of the approach for use at the level of large regions (Siberia, Ural, and etc.), as well as the country level as a whole
- Development of work on the application of approaches for the interpretative capabilities of the obtained model results

---

[26] SHAP. Retrieved from: https://shap.readthedocs.io/en/latest/index.html (Date of access: 08.02.2023)

## 5. Conclusions

The main task of this paper was the construction of an explanatory model for inflation dynamics based on unstructured data from VK on the example of the Omsk region. To be more precise we studied inflation expectations. This goal was complex and to achieve it we performed the following steps:
- We applied some empirical criteria in order to determine the original Omsk group based on prevailing users from the Omsk region. So, we used only groups with more than 2 000 users and in which the share of Omsk residents was at least 20%. There were 1 161 such groups found.
- Then we gathered the corpus of unstructured data consisting of different posts from these target groups. The collected database consisted of 8.5 million records.
- We used RuBERT-tiny (state-of-art model) to build 4 variants of models depending on the maximum length of sentences. Based on the f1-score, as well as the time resources needed to train models, BERT-inflation-256 was chosen as the best model. This model demonstrated better results than the benchmark models (logistic regression, decision tree classifier, random forest classifier, and gradient boosting classifier).
- We applied SHAP for an interpretation of this model. It makes possible to define pro-inflationary and disinflationary types of keywords in different contexts.

In our opinion, this study makes a significant contribution in the direction of inflation analysis based on inflation expectations. We showed the way we could interpret regional pro-inflationary and disinflationary processes with the help of the VK data. As was mentioned before in the macroeconomic sphere the availability of timely and relevant information plays a crucial role, especially while implementing monetary policy within the inflation targeting regime. This methodology allows us to analyze regional inflation expectations with the help of information received from the social network in real time or with a slight delay. At the next step, such data can be used for inflation nowcasting for both country and regional levels.